\documentclass[referee]{mn2e}
\newcommand{\nab}{\mbox{\boldmath $\nabla$}}
\newcommand{\tim}{\mbox{\boldmath $\times$}}
\newcommand{\Om}{\mbox{\boldmath $\Omega$}_{\rm bin}}
\newcommand{\thet}{\mbox{\boldmath $\theta$}}
\newcommand{\rin}{\mbox{$R_{\rm in}$}}

\newcommand{\rmin}{\mbox{$r_{\rm min}$}}

\usepackage{graphicx}
\usepackage{amsmath}
\usepackage{caption}
\usepackage{float}
\usepackage{color}
\graphicspath{{figures/}}

\begin{document}

\title[Disc model for the variability of CoRoT~223992193]{A circumbinary disc
  model for the  
  variability of the eclipsing binary CoRoT~223992193}

\author[C. Terquem, P.~M. S\o rensen--Clark, J. Bouvier] { Caroline
  Terquem$^{1,2}$\thanks{E-mail: caroline.terquem@physics.ox.ac.uk,
    paulmag91@gmail.com, jerome.bouvier@obs.ujf-grenoble.fr}, Paul
  Magnus S\o rensen--Clark$^{3,4}$\footnotemark[1] and J\'er\^ome
  Bouvier$^{3,5}$\footnotemark[1]
  \\
  $^{1}$ Physics Department,
  University of Oxford, Keble Road, Oxford OX1 3RH, UK \\
  $^{2}$ Institut d'Astrophysique de Paris, UPMC Univ Paris 06, CNRS,
  UMR7095, 98 bis bd Arago, F-75014, Paris, France \\
  $^{3}$ Universit\'e Grenoble Alpes, IPAG, F--38000 Grenoble, France
  \\
$^{4}$ I​nstitute of Theoretical Astrophysics, University of Oslo, Oslo 0315,
Norway​ \\
  $^{5}$ CNRS, IPAG, F--38000 Grenoble, France }

\date{}

\pagerange{\pageref{firstpage}--\pageref{lastpage}} \pubyear{}

\maketitle

\label{firstpage}

%
%===========================================================
%

\begin{abstract}
  We calculate the flux received from a binary system obscured by a
  circumbinary disc.  The disc is modelled using two dimensional
  hydrodynamical simulations, and the vertical structure is derived by
  assuming it is isothermal. The gravitational torque from the binary
  creates a cavity in the disc's inner parts.  If the line of sight
  along which the system is observed has a high inclination $I$, it
  intersects the disc and some absorption is produced.  As the system
  is not axisymmetric, the resulting light curve displays variability.
  We calculate the absorption and produce light curves for different
  values of the dust disc aspect ratio $H/r$ and mass of dust in the
  cavity $M_{\rm dust}$.  This model is applied to the high
  inclination ($I=85\degr$) eclipsing binary CoRoT~223992193, which
  shows 5--10\% residual photometric variability after the eclipses
  and a spot model are subtracted.  We find that such variations for
  $I \sim 85\degr$ can be obtained for $H/r=10^{-3}$ and
  $M_{\rm dust} \ge 10^{-12}$~M$_{\sun}$.  For higher $H/r$, $M_{\rm
    dust}$ would have to be close to this lower value and $I$
  somewhat less than $85\degr$.  Our results show that such
  variability in a system where the stars are at least 90\% visible at
  all phases can be obtained only if absorption is produced by dust
  located inside the cavity.  If absorption is dominated by the parts
  of the disc located close to or beyond the edge of the cavity, the
  stars are significantly obscured.
\end{abstract}


%
%===========================================================
%

\section{Introduction}
\label{sec:intro}
%-----------------

In the course of CSI~2264, an international photometric monitoring
campaign of several hundred young stars located in the NGC2264 star
forming region (Cody et al. 2014), Gillen et al. (2014) reported the
discovery of CoRoT~223992193, a double--lined eclipsing binary
consisting of two M--type pre--main sequence (PMS) stars. The analysis of
the CoRoT~223992193 light curves completed by radial velocity measurements
yielded a complete characterisation of the orbit of the tight system
(P=3.8745745~days) as well as a precise determination of the mass and
radius of each component.  At an age of 3.5--6 Myr, this eclipsing
system provides strong constraints on the validity of low--mass PMS
evolutionary models (Gillen et al. 2014, Stassun et al. 2014).

Interestingly, the CoRoT~223992193 light curve also reveals
significant variability outside of the eclipses.  Part of the large
amplitude (10--15\%) and rapidly evolving out--of--eclipse variations
can be accounted for by surface spots, but there is a residual
variability with an amplitude of 5--10\% (Gillen et al. 2015).  Based
on the high system's inclination ($I$=85$\degr$) and the presence of
significant near/mid--infrared excess, it was suggested that at least part
of the observed variability could be due to hot dust orbiting the inner
system and, as it rotates, partially occults the central stellar
components (Gillen et al. 2014, 2015).
%Indeed, the out-of-eclipse variations were found to be periodic, with
%a period close to, but different from, that of the stellar eclipses
%(P$_{ooe}= 3.644$~days).

Here, we investigate this possibility further by running numerical
simulations of a binary system, whose properties are those of
CoRoT~223992193, and which is surrounded by a circumbinary dusty disc.
Similar models of discs around binaries have been published previously
(e.g., Hanawa et al. 2010, de~Val--Borro et al. ~2011, Shi et
al.~2012, Shi \& Krolik~2015), but our numerical modelling
specifically aims to reproduce the out--of--eclipse variability of
CoRoT 223992193.   Assuming that the system is seen along a line of
sight with a given inclination angle $I$, we  calculate the mass
of dust that the line of sight intersects from the disc, and the
resulting light curve.   

The plan of the paper is as follows.  In section~\ref{sec:numeric}, we
present two dimensional  numerical simulations of the circumbinary
disc around a binary system like CoRoT~223992193.  In
section~\ref{sec:lightcurve}, we explain how the light curves are
calculated and derive the parameters that give the best fit to the
CoRoT~223992193 system.  Finally, in section~\ref{sec:discussion}, we
summarise and discuss our results.

%
%===========================================================
%

\section{Numerical simulations of circumbinary discs}
\label{sec:numeric}
%-----------------

We consider a binary system and a circumbinary disc which lies in the
orbital plane.  The angular velocity of the binary is
$\Omega_{\rm bin} = \sqrt{G(M_1+M_2)/a^3}$, where $G$ is the
gravitational constant, $M_1$ and $M_2$ are the masses of the stars
and $a$ is the separation.  The binary is assumed to have a circular
orbit.  In the frame corotating with the binary and with origin at the
centre of mass, in two dimensions, the equation of motion for a fluid
element in the disc with position vector ${\bf r}$ and velocity
${\bf v}$ is:

\begin{equation}
\frac{\partial {\bf v}}{\partial t} + \left( {\bf v} \cdot \nab
\right) {\bf v} = -\frac{1}{\Sigma} \nab P - \nab \Phi - \Om \tim
\left( \Om \tim {\bf r} \right) - 2 \Om \tim {\bf v},
\label{eom}
\end{equation}
where $P$ is the pressure averaged over the disc scale height,
$\Sigma$ is the surface mass density,
$\Om = \Omega_{\rm bin} \hat{\bf z}$, with $\hat{\bf z}$ being the
unit vector perpendicular to the disc plane, and $\Phi$ is the
gravitational potential due to the stars.  We assume that the disc's
mass is small compared to that of the stars, so that self--gravity can
be neglected.  We have:
\begin{equation}
\Phi= - \frac{G M_1}{\left| {\bf r} - {\bf r}_1 \right|} - \frac{G
  M_2}{\left| {\bf r} - {\bf r}_2 \right|} ,
\label{Phi}
\end{equation}
where ${\bf r}_1$ and ${\bf r}_2$ are the position vectors of the
stars with masses $M_1$ and $M_2$, respectively. 

The surface mass density also has to satisfy the mass
conservation equation:
\begin{equation}
\frac{\partial {\Sigma}}{\partial t} + \nab \cdot \left( \Sigma {\bf
    v} \right) = 0.
\label{cons}
\end{equation}

Finally, we adopt an isothermal equation of state for the
gas:
\begin{equation}
P=\Sigma c_s^2,
\label{eos}
\end{equation}
where $c_s$ is the isothermal sound speed.  

\subsection{Initial conditions}

In the corotating frame with origin at the centre of mass, we note
$(x,y,z)$ the cartesian coordinates, where $x$ is the axis passing
through the centres of the stars and $z$ is perpendicular to the
orbital plane, and $(r, \theta, z)$ the associated polar coordinates.
The coordinate of the primary star, of mass $M_1$, is
$x_1=-a M_2/ (M_1+M_2)$, whereas the coordinate of the secondary star,
of mass $M_2$, is $x_2=a M_1/ (M_1+M_2)$.  Initially, the disc extends
from an inner radius $\rin$ to some outer radius.

Following Hanawa et al. (2010), we take the initial density
profile to be:
\begin{equation} 
\Sigma_{\rm init}=\Sigma_0 \left( 0.55+0.45 \tanh \frac{r- \rin }{h}
  \right)  +\Sigma_{\rm min}  .
\label{sigmainit}
\end{equation}
The constant $\Sigma_0$ is related to the total mass of the disc and
$\Sigma_{\rm min}$, which we choose to be very small compared to
$\Sigma_0$, is added to ensure that the mass density does not become
smaller than some threshold to avoid numerical problems.  The above
equation is such that $\Sigma_{\rm init} \simeq 0$ at the disc's inner edge,
increases to $\Sigma_0$ over a scale $h$ and is uniform equal to
$\Sigma_0$ beyond.

\noindent As the binary perturbs significantly the disc only over a
very limited region near its inner edge, over which the surface
density is not expected to vary much, we will only consider models
with an initial uniform surface density, as given by
equation~(\ref{sigmainit}).

The initial velocity in the disc is set to be the Keplerian
velocity in the rotating frame:
\begin{equation}
{\bf v}_{\rm init} = r \left( \Omega_{\rm K} -  \Omega_{\rm bin}
\right) \hat{\thet} ,
\label{vinit}
\end{equation}
with $\Omega_{\rm K} = \sqrt{G(M_1 +M_2)/ r^3}$ and $\hat{\thet}$ is
the unit vector in the azimuthal direction. 

\subsection{Units}

In the numerical simulations, we take $G=1$, $a=1$, $M_1=0.7$ and
$M_2=0.5$.  Therefore, all the lengths will be given in units of the
binary separation.  In this unit system, the timescale
$\Omega_{\rm bin}^{-1}$ is approximately $0.91$ and the orbital period is
$ 2 \pi /\Omega_{\rm bin} \simeq 5.74$.  

\noindent With the equation of state~(\ref{eos}), the pressure term in
equation~(\ref{eom}) does not depend on $\Sigma_0$.  Since, in
addition, the disc is non self--gravitating, its mass does not enter
the equations.  Therefore we can take $\Sigma_0=1$.

\subsection{Numerical set up and boundary conditions}

Equations~(\ref{eom}), (\ref{cons}) and~(\ref{eos}) are solved with
the initial conditions~(\ref{sigmainit}) and~(\ref{vinit}) using 
version 4.0 of the PLUTO code (Mignone et al. 2007).  No explicit
viscosity is used. 

Although a polar coordinate system would be a natural choice to
describe the evolution of the disc, it would involve a singularity at
the origin and therefore would not be numerically tractable.
Therefore, we set up a cartesian grid.  We use three consecutive
adjacent grids in the $x$ and $y$ directions.  They extend from $-15$
to $-2$, $-2$ to $2$ and $2$ to $15$ and each has 1024 zones in both
directions.  The outer radius of the disc does not need to be
specified as long as it is beyond the edge of the grid.

To prevent the gravitational potential from becoming
infinite when a particle approaches one of the stars very closely, we
soften it by replacing $\left| {\bf r} - {\bf r}_{1,2} \right|$ by
${\rm max} \left( \left| {\bf r} - {\bf r}_{1,2} \right| , d_{\rm
    soft} \right)$
in equation~(\ref{Phi}), where $d_{\rm soft}$ is the softening
parameter.  We take $d_{\rm soft}=0.07$.

When the surface density becomes very small, the code does not perform
very well.  Therefore, we fix a threshold $\Sigma_{\rm min}=10^{-3}$
and replace the density by this value wherever it becomes smaller.
Following Hanawa et al. (2010), we also fix $h=0.144$ in
equation~(\ref{sigmainit}).

To prevent wave reflection at the outer edge of the grid, we use the
wave killing condition described in de~Val--Borro et al. (2006).  This
procedure forces the variables to relax toward their initial values on
some prescribed timescale in a region close to the outer edge of the
grid.

The wave killing procedure is used when waves would otherwise be
reflected at the grid outer boundary.  Therefore, this boundary stays
close to its initial state and has little effect on the dynamics of
the flow in the disc's inner parts.  Therefore, our boundary
conditions are:
\begin{equation}
F=F_{\rm init},
\end{equation}
at the edge of the grid, where $F$ is either $\rho$, $P$, $v_x$ or $v_y$.

\subsection{Results}

\begin{figure}
\begin{center}
{\includegraphics[scale=0.3]{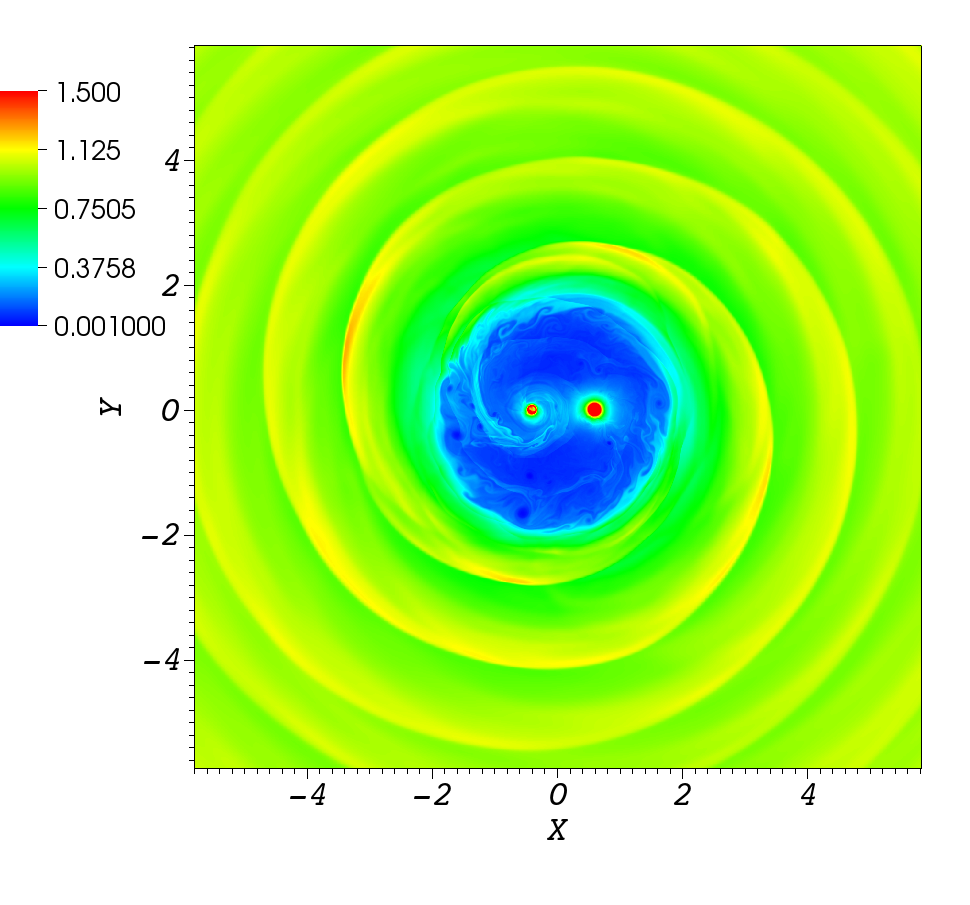}}
\caption{Pseudocolor plot of the surface density $\Sigma$ in
  the $x$--$y$ plane for a sound speed $c_s=0.3$.  The primary and
  secondary stars are located at $(-0.42,0)$ and $(0.58,0)$,
  respectively. The mapping of data to color values is linear.  }
\label{fig1}
\end{center}
\end{figure}

Figure~\ref{fig1} shows the surface density $\Sigma$ in the $x$--$y$
plane in a disc that has reached a steady--state for a sound speed
$c_s=0.3$.   \\

\noindent The circumbinary disc is set up with an inner edge at
$\rin=2.5$.  As can be seen from figure~\ref{fig1}, the material near
this edge is pushed inward by pressure forces.  The gravitational
torque exerted by the binary opposes pressure forces, so that a cavity
with a radius of about~2 is created.  This is consistent with theory,
which predicts a disc's inner radius of roughly twice the binary
separation (Lin \& Papaloizou 1979).  Some mass though does penetrate
inside the cavity, as seen in previous simulations.

\noindent The maximum value of $\Sigma$ along the density waves
launched at the inner edge of the disc is about 1.5.  Circumstellar
discs form around the stars.  Their radius is about 0.3--0.4 for the
primary star and 0.2--0.3 for the secondary star (as seen in
figure~\ref{fig1}, the disc around the secondary is denser but
actually not as extended).  Again, this is consistent with theory, as
tidal truncation of the circumstellar discs by the other star is
expected to limit their outer radii to about one third of the binary
separation (Paczynski~1977, Papaloizou \& Pringle~1977).  In the inner
parts of these circumstellar discs, $\Sigma$ becomes large as material
piles up.  In figure~\ref{fig1}, $\Sigma$ in the inner parts of the
circumstellar discs is saturated.  In reality though, mass from these
discs would be accreted onto the stars and/or launched into jets.  As
we will see below, since dust in the inner parts of the circumstellar
discs is sublimated by  stellar radiation, the detailed structure
of these discs does not affect the results presented in this paper.

As this map of the surface density around a binary system is very
typical, and is similar to that obtained in many previous studies
(see, e.g., Hanawa et al. 2010, Shi \& Krolik 2015), we will focus on
this single particular case in this paper.  We have run models with
different sound speeds but they all give similar results.

\section{Binary light curve}
\label{sec:lightcurve}
%-----------------

\subsection{Disc vertical structure}

The numerical simulations we have performed are two dimensional, and
therefore we only calculate the surface mass density:
\begin{equation}
\Sigma (x,y) = \int_{-\infty}^{\infty} \rho(x,y,z) dz,
\label{sigmaint}
\end{equation}
where $\rho$ is the density per unit volume.  To generate the third
dimension, we assume that the disc is isothermal in the vertical
direction with a scale--height $H$, so that:
\begin{equation}
\rho(x,y,z) = \rho_0(x,y) \;  {\rm e}^{- z^2 / (2 H^2) },
\label{rho3D}
\end{equation}
where $\rho_0$ is the density in the disc midplane (at $z=0$).  Then
equation~(\ref{sigmaint}) yields:
\begin{equation}
\Sigma (x,y) = \sqrt{2 \pi} \rho_0(x,y) H(x,y),
\end{equation}
where $H$ may depend on $x$ and $y$.  
%Thereafter, we will take $H=H_0r$, where $H_0$ is a constant and
%$r=\sqrt{x^2+y^2}$ is the radius in cylindrical coordinates.
Then equation~(\ref{rho3D}) becomes:
\begin{equation}
\rho(x,y,z) = \frac{\overline{\rho} \Sigma (x,y)}{\sqrt{2 \pi} H}  \;  {\rm e}^{- z^2 / (2
  H^2) },
\label{rho3D2}
\end{equation}
where $\overline{\rho}$ is a constant factor by which we scale the
density obtained in the simulations (it can be chosen by requiring a
given total mass for the disc).  Equation~(\ref{rho3D2}) is used to
construct $\rho$ from $\Sigma$.

The simulations presented in section~\ref{sec:numeric} were done for a
constant sound speed $c_s=0.3$ in code units.  When hydrostatic
equilibrium is reached in the vertical direction, this corresponds to
a gas disc scale height $H=c_s/\Omega_{\rm K}=0.27 r^{3/2}$.  For the
calculation to be self--consistent, this is the value that should be
used when calculating the binary light curve.  However, the gas is
completely transparent to stellar radiation, and absorption is due
solely to {\em dust} grains which are mixed with the gas.  The dust
disc scale height might differ significantly from the gas scale height
if, for example, dust grains are not well mixed all the way up with
the gas. Therefore, in the following section, we will allow $H$ to
depart from the value given above.  For simplicity, we will consider
$H/r={\rm const}$, and will vary this value until we obtain a light
curve which matches that of the system which is being modelled.  The
implications of the value of $H/r$ found below will be discussed in
section~\ref{sec:discussion}.

\subsection{Light curve modelling}

\begin{figure}
\begin{center}
{\includegraphics[scale=0.3]{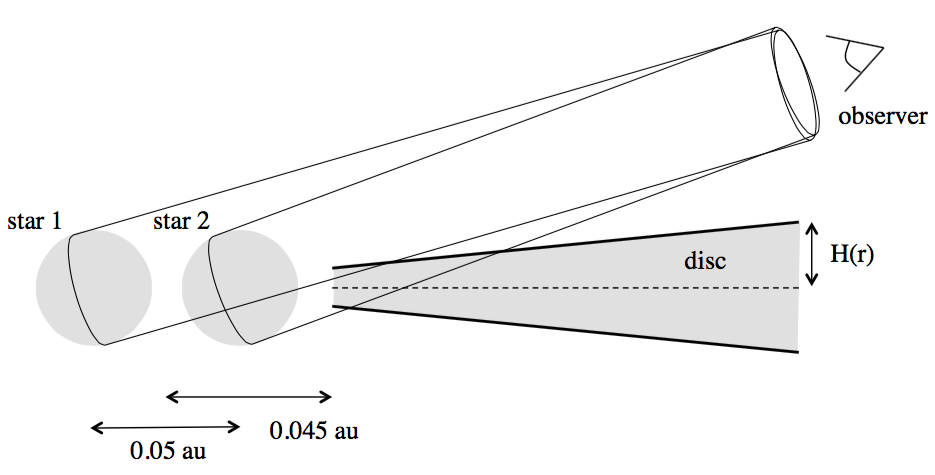}}
\caption{Schematic view of the system for which we calculate
  light curves.}
\label{fig2}
\end{center}
\end{figure}

To obtain the light curve of the system, we calculate, for each star,
the amount of energy that comes out of a cylinder whose base is the
projected surface of the star and whose axis is the line of sight (see
figure~\ref{fig2}).  Part of the flux emitted by the stars is absorbed
by the disc if the cylinders intersect it.  Each cylinder is split
into multiple line of sights in the z--direction and sliced up into a
number of bins $j=1, \dotsc, n$ along each line of sight.  The total
mass $m_j$ in each bin is the mass in the part of the disc which is
intersected by the bin.  To obtain this mass, we integrate each
density datapoint over its volume element, assign a weight to the
volume elements in the disc depending on how far they are from the
axis of the cylinder, and sum up over all these points.  We 
calculate a mean density $\rho_j$ in the bin by dividing the mass
$m_j$ by the volume of the bin.  The intensity transferred from the
bin $j$ to the bin $j+1$ (going away from the star) is therefore
$I_{j+1}=I_j {\rm e}^{-\tau_j}$, where
$\tau_j= \kappa \rho_j \Delta l_j$ is the optical depth, with $\kappa$
being the opacity and $\Delta l_j$ being the length of the bin along
the axis of the cylinder.  In the calculations presented below, we
will take $\kappa=5$~cm$^2$~g$^{-1}$ (Jensen \& Mathieu~1997).  This
assumes that gas and dust are well mixed and that the gas to dust
ratio is 100.

In this paper, we calculate the light curve of a system which has the
characteristics of CoRoT~223992193 (Gillen et al. 2014).  The primary
star has a mass $M_1 =0.7$~M$_\odot$ and a radius $R_1=1.3$~R$_\odot$,
the secondary has a mass $M_2 =0.5$~M$_\odot$ and a radius
$R_2=1.1$~R$_\odot$ and their separation is $a=0.05$~au.  So the unit
of distance $a$ that we have used in the previous section corresponds
to a physical distance of 0.05~au.  Using the effective temperatures
$T_1=3700$~K and $T_2=3600$~K estimated by Gillen et al. (2014), we
infer the luminosities $L_1=0.28$~L$_\odot$ and $L_2=0.18$~L$_\odot$
(which corresponds to $L_2/L_1=0.64$, as observed).

\noindent The sublimation radii are 3.9 and 3.2 solar radii for the
primary and secondary stars, respectively (Gillen et al. 2014).  It
means that no obscuration of the stars can come from material within a
distance of 0.37 and 0.30 from the primary and secondary stars,
respectively.  To calculate the emergent stellar intensities, we
therefore exclude the parts of the disc within
$\rmin = {\rm max}\left( 0.42+0.37, 0.58+0.30 \right) \simeq
0.9=0.045$~au
(the primary and secondary stars are located at $x=-0.42$ and
$x=0.58$, respectively).

\noindent In the simulations presented in the previous section, the
disc is only modelled up to some outer radius given by the boundary of
the grid.  For the range of line of sight inclinations we consider, we
find that absorption of the stellar fluxes is due to the parts of the
disc which are much smaller than this outer radius.  Therefore the
finite size of the grid is not a limitation of the model.

%\noindent 
%The results presented below correspond to a disc with a semithickness
%$H$ such that $H/r=H_0={\rm const}$.

\noindent Figure~\ref{fig2} gives a schematic view of the system.  In
this sketch, the disc has been represented as though it had a finite
thickness, but there is actually mass at all altitudes above the
midplane as the density varies according to equation~(\ref{rho3D}).
Therefore, there is always some mass in every bin along the cylinders,
even if only a tiny amount.  \\

We define the viewing angle $\theta$ as the angle between the
projection of the line of sight in the $(x,y)$ plane and the
$x$--axis.  A viewing angle $\theta=0$ means that the observer looks
at the system along the line joining the centres of the stars and with
the secondary being in front (primary eclipse), i.e. from the right on
figures~\ref{fig1} and~\ref{fig2}.  This angle increases when the line
of sight rotates clockwise for a fixed binary, as in
figure~\ref{fig1}, which is equivalent to rotating the binary
anti--clockwise for a fixed line of sight (as in the realistic
situation).  We also define the inclination angle $I$ between the line
of sight and the perpendicular to the orbital plane ($I=0$ means that
the system is seen pole--on).

The light curve is the amount of flux received from the system as a
function of time for a fixed line of sight.  If the disc is in 
quasi--steady state, as is the case here, rotating the disc for a
fixed line of sight is equivalent to varying the viewing angle
$\theta$.  Therefore,  the light curve of the system is obtained by
calculating the flux as a function of $\theta$.  

\subsection{Results}

As mentioned in the introduction, the light curve of CoRoT~223992193
displays out--of--eclipse variations which have a peak--to--peak
amplitude of about 15\% (Gillen et al 2014).  Part of this variability can be explained by
a spot model, but the residuals show extra variability with an
amplitude of about 5-10\% (Gillen et al. 2015).   
The system is
seen with $I=85\degr$ and the visual extinction is estimated to be
between 0 and 0.1
(Gillen et al. 2014), which means that there could be an extinction of the
stars of up to 10\% at all orbital phases.   

Here, we
investigate whether the residual variability can 
be explained by obscuration of the stars by the circumbinary and/or
circumstellar discs.

We calculate the light curve of the system in the way described in the
previous section, and vary $H/r$ and $\overline{\rho}$ (see
eq.~[\ref{rho3D2}]) until we obtain a curve which displays 5--10\%
variations for values of $I$ close to 85$\degr$.  We define the
disc's inner cavity as the region between $r=\rmin=0.9$ and $r=2$.
We present results
for $H/r=10^{-3}$ and 0.05, and for values of $\overline{\rho}$ corresponding
to a mass of {\em dust} in the disc's inner cavity 
$M_{\rm dust} =10^{-12}$ and $10^{-11}$~M$_{\sun}$.

%We present results for $H/r=10^{-3}$ and
%$\overline{\rho}=1.9 \times 10^{-10}$~g~cm$^{-3}$ or
%$\overline{\rho}=1.9 \times 10^{-9}$~g~cm$^{-3}$, and for $H/r=0.05$
%and $\overline{\rho}=3.9 \times 10^{-12}$~g~cm$^{-3}$ or
%$\overline{\rho}=3.9 \times 10^{-11}$~g~cm$^{-3}$.  For each value of
%$H/r$, the lowest $\overline{\rho}$ corresponds to a mass of {\em
%  dust} in the disc's inner cavity of
%$M_{\rm dust} = 10^{-12}$~M$_{\sun}$, whereas the largest
%$\overline{\rho}$ corresponds to $M_{\rm dust} = 10^{-11}$~M$_{\sun}$.

\noindent The largest value of $H/r$ we consider would be expected in
a disc in which the sound speed is about 10\% of the Keplerian
velocity and the dust is well mixed with the gas all the way through
the disc's thickness, whereas the smallest value would be typical of a
disc with very small pressure forces and/or dust not well mixed with
the gas.  Note that in this calculation we assume that the mass of
dust is about one hundredth of the mass of gas, and the two components
are well mixed.  This assumption will be discussed in
section~\ref{sec:discussion}.

\begin{figure}
\begin{center}
{\includegraphics[scale = 0.5]{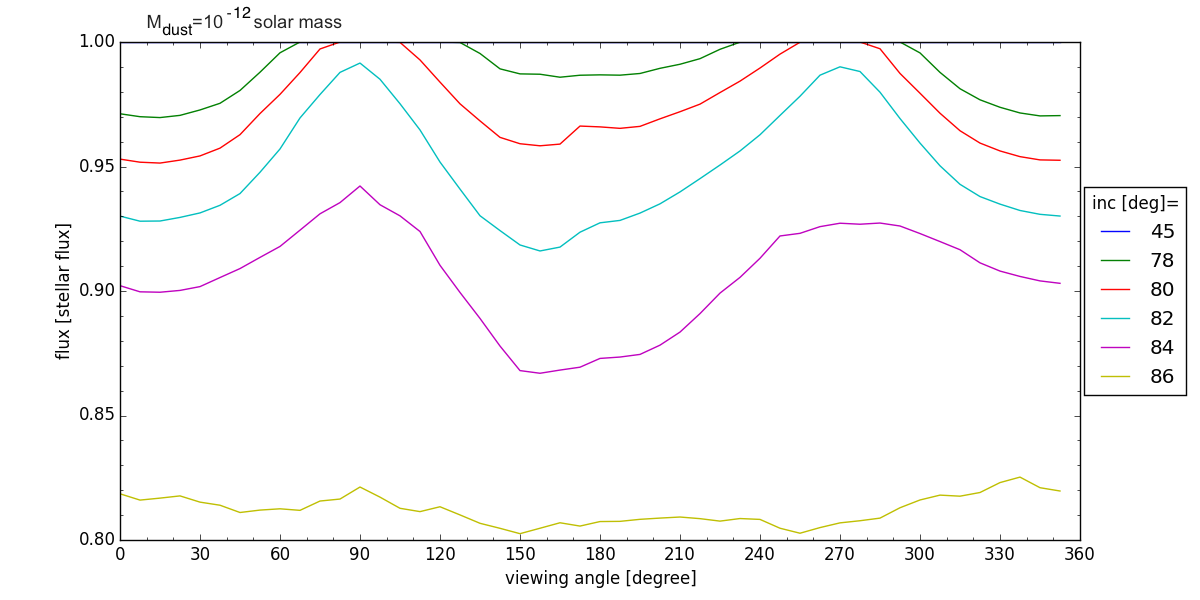}} \\
{\includegraphics[scale  = 0.5]{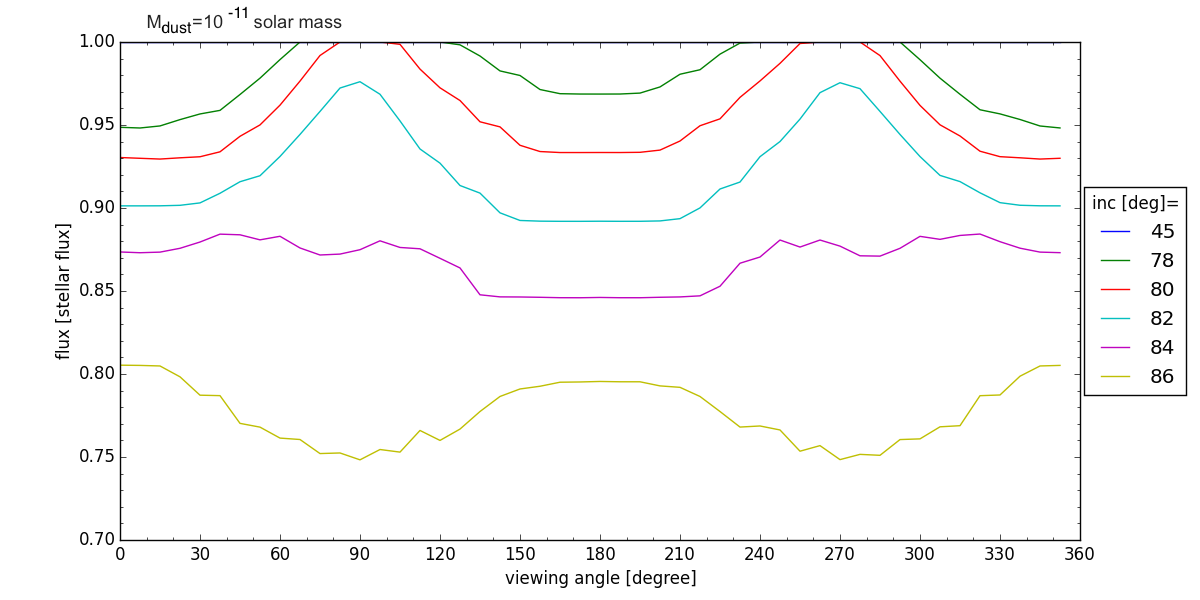}} 
\caption{Flux divided by the unobscured value {\em versus} viewing
  angle (in degrees) for $H/r=10^{-3}$.  The upper and lower plots
  correspond to $M_{\rm dust} = 10^{-12}$ and $10^{-11}$~M$_{\sun}$,
  respectively.  The curves have been calculated for different
  inclination angles $I=45$, 78, 80, 82, 84 and 86$\degr$.  The curve
  corresponding to $I=45\degr$ is flat and corresponds to a flux equal
  to~1.  The curves which ressemble most that of CoRoT~223992193 are
  the one corresponding to $I=78$--$82\degr$.  In the case of
  $M_{\rm dust} = 10^{-12}$~M$_{\sun}$, $I=84\degr$ would also give a
  good fit.  }
\label{fig3}
\end{center}
\end{figure}

\begin{figure}
\begin{center}
{\includegraphics[scale = 0.5]{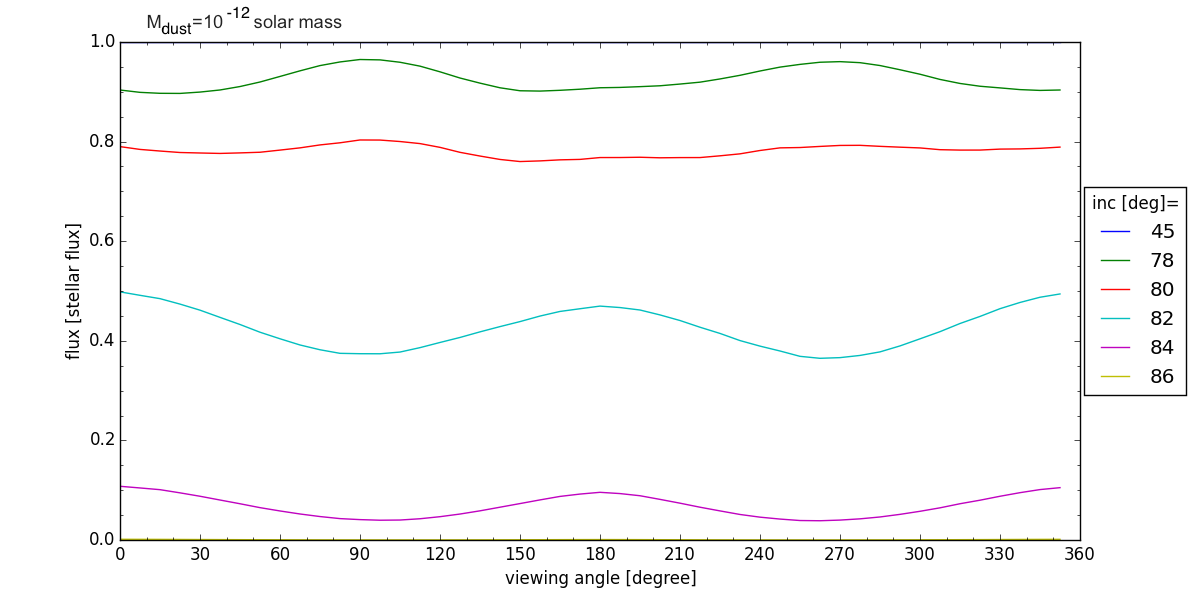}} \\
{\includegraphics[scale  = 0.5]{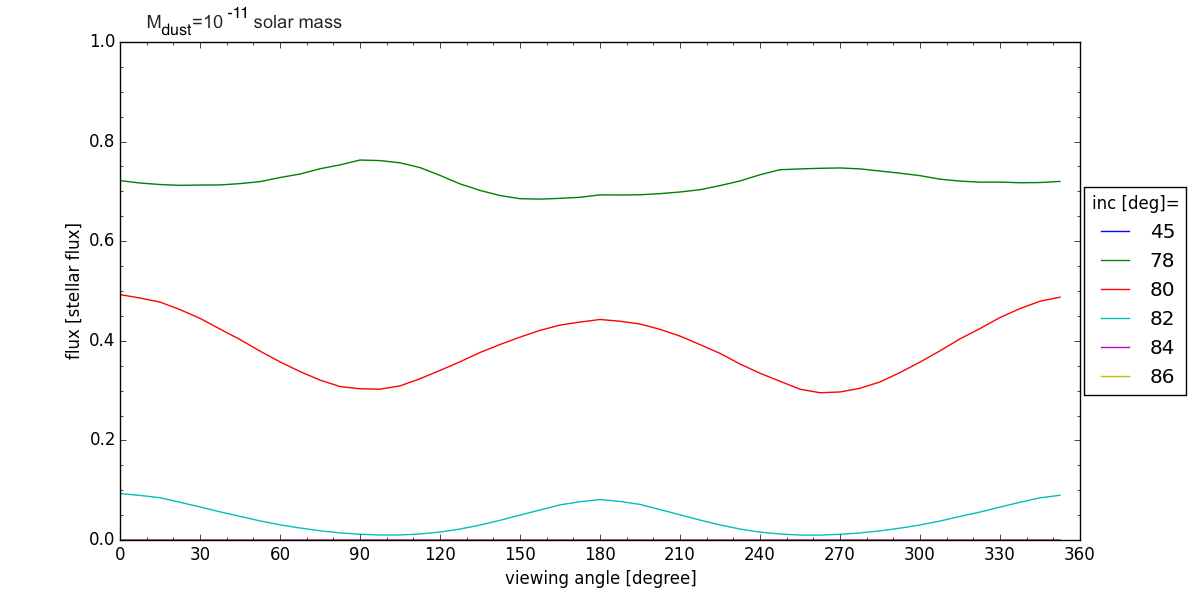}} 
\caption{Same as figure~\ref{fig3} but for $H/r=0.05$.  
Here, only $I=78\degr$ for  $M_{\rm dust} = 10^{-12}$~M$_{\sun}$ would
match the CoRoT~223992193 light curve.
}
\label{fig4}
\end{center}
\end{figure}

\noindent The flux {\em versus} viewing angle is displayed in
figure~\ref{fig3} for $H/r=10^{-3}$ and $M_{\rm dust} = 10^{-12}$ and
$10^{-11}$~M$_{\sun}$, and in figure~\ref{fig4} for $H/r=0.05$ and the
same values of $M_{\rm dust}$.  Note that the flux has been divided by
the value it would have if the stars were unobscured (i.e. if there
were no disc).

\noindent The eclipses of the stars by each other have not been taken
into account in the calculation of the light curves, i.e. the emergent
stellar fluxes have been calculated separately for the two cylinders
represented in figure~\ref{fig2} and then added.  The curves
therefore show the out--of--eclipse variations.  The eclipses would
produce additional narrow ``dips'' in the light curves at $\theta=0$
and 180$\degr$.

\noindent When $\theta=0$, the cylinder based on the secondary star
intersects a larger portion of the disc than the cylinder based on the
primary star (see figure~\ref{fig2}).  Therefore, the secondary star
is at least partially hidden for the inclinations $I$ we have
considered (except 45$\degr$), while the primary may still be fully
visible.  If the secondary star is completely hidden and the primary
star fully visible, the emergent flux divided by the unobscured value
is $L_1/(L_1+L_2)=0.6$.  We can see that, for $H/r=0.05$, the flux
becomes smaller than 0.6 for the largest values of $I$, indicating
that the primary star is also partially hidden.  For
$\theta=180\degr$, it is the same situation except that it is now the
primary star which is at least partially hidden.  When $\theta$ is
around 90 or 270$\degr$, the stars are further away from the edge of
the disc located at $r=\rmin$, and therefore the cylinders intersect
less mass from the disc than when $\theta=0$ or 180$\degr$.  As a
result, both stars are (almost) fully visible (within 10\%) for
$H/r=10^{-3}$ and $I \le 80$--$82\degr$ for both values of
$M_{\rm dust}$ considered.  For $H/r=0.05$, this only happens for
$I=78\degr$ and the lowest value of $M_{\rm dust}$.

\noindent We have checked that, when $M_{\rm dust}$ is further
increased above $10^{-11}$~M$_{\sun}$, the curves corresponding to
$H/r=10^{-3}$ do not change (they stay the same as for
$M_{\rm dust}=10^{-11}$~M$_{\sun}$), whereas for $H/r=0.05$ the flux
becomes smaller.

Figure~\ref{fig3} shows that variations with an amplitude on the order
of 5\%--10\% for $I$ close to $85\degr$ can be obtained for
$H/r=10^{-3}$ and a mass of dust in the cavity above
$\sim 10^{-12}$~M$_{\sun}$.  Figure~\ref{fig4} indicates that such
variations can also be obtained for $H/r=0.05$, but only for the
smallest values of $M_{\rm dust}$ and for rather low values of the
inclination angles $I$.

Note that Gillen et al. (2014) had estimated that a mass of dust of at
least $10^{-13}$~M$_{\sun}$ was needed in the disc inner cavity to produce
the observed out--of--eclipse variations of CoRoT~223992193.

\begin{figure}
\begin{center}
{\includegraphics[scale=0.5]{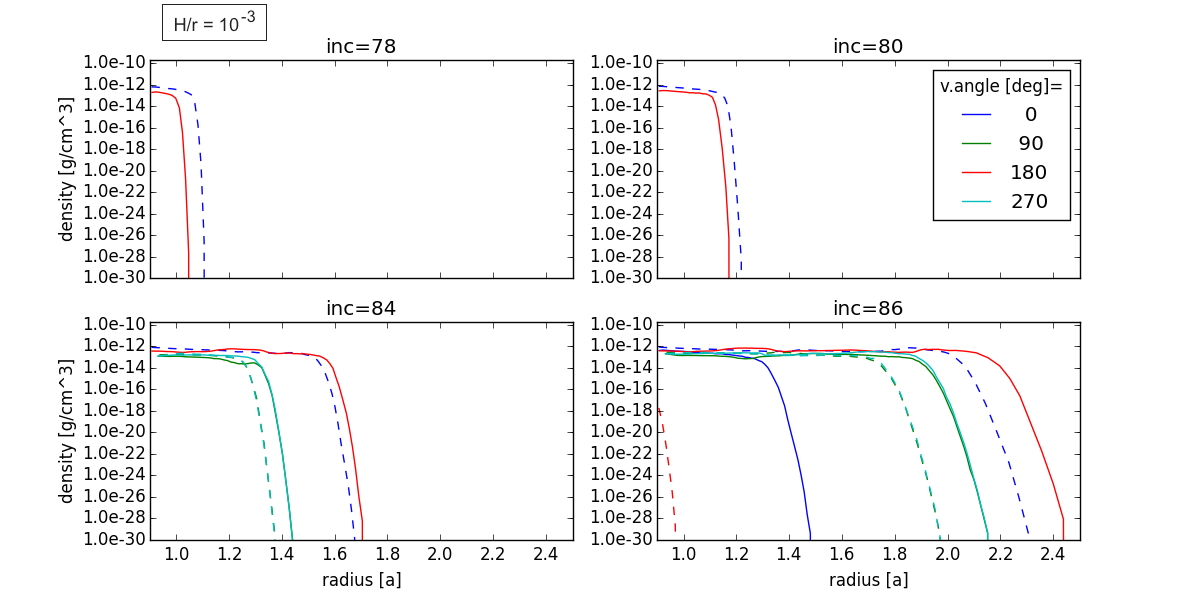}} \\
{\includegraphics[scale=0.5]{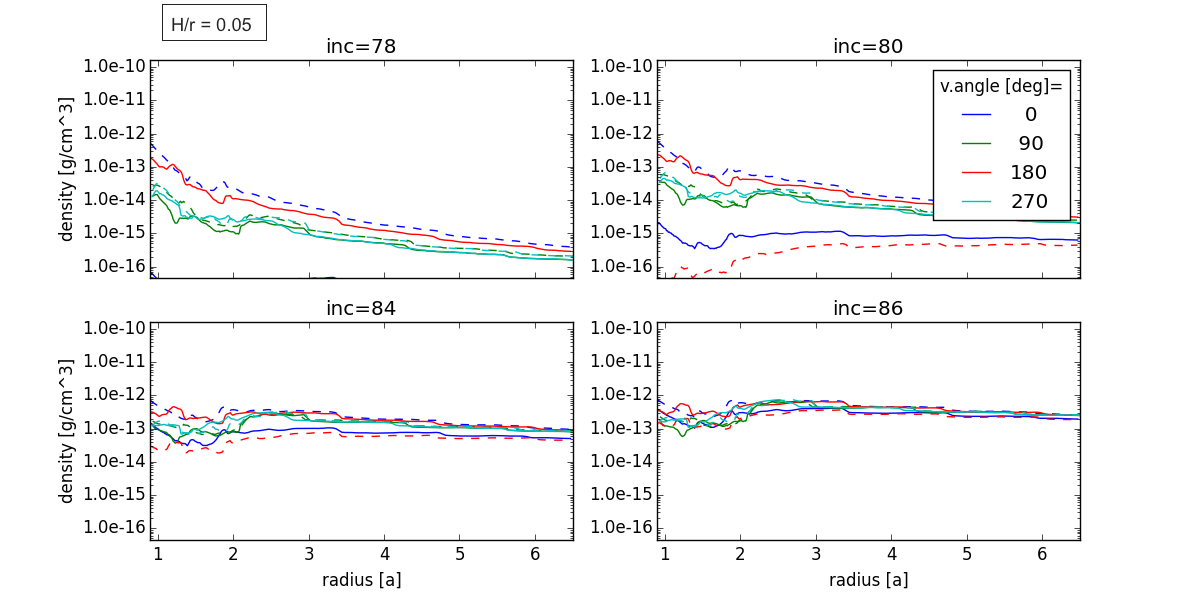}}
\caption{Mass density of gas (in g~cm$^{-3}$ and in logarithmic scale) along
  the line of sight toward the primary (solid lines) and secondary
  (dashed lines) stars {\em versus} radius $r$ in the disc (in units
  of the binary separation $a$) for
  $M_{\rm dust} = 10^{-12}$~M$_{\sun}$.  The four upper and four lower
  plots correspond to $H/r=10^{-3}$ and $H/r=0.05$, respectively.  For
  each value of $H/r$, the four different plots correspond to $I=78$,
  $80$, $84$ and $86\degr$.  In each case, the calculations are done
  for the viewing angles $\theta=0, 90, 180$ and $270\degr$.  The
  corresponding curve is not shown when the density is too small.  For
  the values of $H/r$ and $I$ for which the variations of the flux
  ressemble that of CoRoT~223992193, absorption is always produced by
  material located inside the disc's inner cavity.  }
\label{fig5}
\end{center}
\end{figure}

To see which parts of the disc contribute most to the absorption of
the stellar fluxes, we show the mass density of gas along the line of sight
toward the stars for $H/r=10^{-3}$ and $H/r=0.05$ and for
$M_{\rm dust} = 10^{-12}$~M$_{\sun}$ in figure~\ref{fig5}.  The density is
plotted as a function of radius $r$ {\em in the disc} (i.e. the
distance along the line of sight projected onto the disc's plane).
The different plots are for different inclination angles $I$,
and in each plot the different curves correspond to different viewing
angles $\theta$.  The solid and dashed lines show absorption
along the line of sight toward the primary and secondary stars,
respectively.  The curves corresponding to $M_{\rm dust} =
10^{-11}$~M$_{\sun}$ are exactly the same except for the density being
10 times larger.

\noindent For the values of the parameters for which the curves
displayed in figures~\ref{fig3} and~\ref{fig4} give a good fit to the
CoRoT~223992193 light curve, absorption is always due to material
inside the disc's inner cavity.

\begin{figure}
\begin{center}
{\includegraphics[scale=0.5]{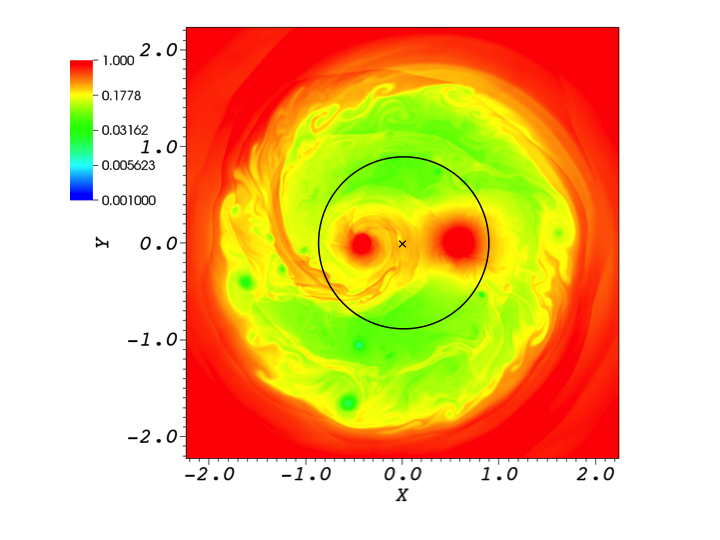}}
\caption{As in figure~\ref{fig1} but with a logarithmic mapping of
  data to color values.  The cross is at the location of the centre of
  mass of the binary and the circle, of radius $r=\rmin=0.9$, shows
  the region which is excluded when calculating the light curves.  }
\label{fig6}
\end{center}
\end{figure}

In figure~\ref{fig6}, we show the same density map as in
figure~\ref{fig1}, but with a zoom on the inner parts of the disc and
with a logarithmic mapping of data to color values.  The circle, of
radius $r=\rmin$, shows the region which is excluded when calculating
the light curves.  As pointed out above, there are no dust grains in
this region.  It is clear from this figure that absorption comes from
inside the circumbinary disc's cavity, probably mainly from the edge
of the circumstellar discs and from the streams of material which are
being accreted by the stars.  Figure~\ref{fig3} shows a ``dip'' in the
light curves near $\theta=150\degr$.  It is likely to be produced by
the stream linking the circumbinary disc to the primary star, as seen
in figure~\ref{fig6}.  Such a dip has been observed in CoRoT~223992193
light curve and does happen before secondary eclipse (Gillen et
al. 2015), as observed here.

\section{Summary and discussion}
\label{sec:discussion}
%-----------------

In this paper, we have calculated the flux received from a binary
system surrounded by a circumbinary disc.  The disc is modelled using
two dimensional hydrodynamical simulations, and the vertical structure
is derived by assuming it is isothermal.  Because of the gravitational torque exerted
by the binary, a cavity is created in the inner parts of the
circumbinary disc.  Apart from in the circumstellar discs around the stars, 
the surface density is much lower in this cavity than in the rest of the disc.
As seen in all previous simulations of such systems, accretion streams
linking the outer edge of the cavity and the circumstellar discs are
present.   

If the line of sight along which the system is observed has a high
inclination $I$, it intersects the disc and some absorption is
produced by the dust in the disc.  As the system is not axisymmetric,
the emergent flux varies depending on the angular coordinate in the
disc along which the system is observed.  We have found that
variations on the order of 5--10\% for $I$ close to 85$\degr$ could be
obtained for a disc's aspect ratio $H/r=10^{-3}$ and a mass of {\rm
  dust} $M_{\rm dust}$ in the cavity above $\sim 10^{-12}$~M$_{\sun}$.
If $H/r=0.05$, we obtain a 5\% variation only for
$M_{\rm dust}= 10^{-12}$~M$_{\sun}$ and a rather low value of
$I=78\degr$.  For higher masses and/or higher values of $I$, the stars
are significantly obscured.   Our results show that
5--10\% variability in a system where the stars are at least 90\%
visible at all phases can be obtained only if absorption is produced
by dust located inside the disc's inner cavity.  If absorption is
dominated by the parts of the disc located further away, close to or
beyond the inner edge of the cavity, the stars are heavily obscured.

Although we have presented results for only one simulation of a
circumbinary disc corresponding to a sound speed $c_s=0.3$, we have
run cases with $c_s=0.2$ and 0.25 and checked that the light curves
were very similar in all cases.  When the sound speed is decreased,
pressure forces are less efficient at smoothing out density
enhancement, and regions of larger density become more extended.
However, as we have found, the type of absorption which is observed in
CoRoT~223992193 is produced in the disc's inner cavity, and therefore
the parameters that give the best fit to this system light curve do no
depend on the disc's sound speed.

To model the gas disc, we have fixed $c_s=0.3$, which corresponds to a
gas scaleheight at hydrostatic equilibrium
$H=c_s/\Omega_{\rm K}=0.27 r^{3/2}$.  However, our best fit for the
CoRoT~223992193 light curve is obtained for a rather low value of
$H/r=10^{-3}$.  Although we have assumed in our calculation of the
light curve that dust and gas were well mixed, the value of $H$ we
constrain is that of the {\em dust} disc, as absorption is due solely
to dust grains.  In reality, the thickness of the gas and dust discs
may be very different.  Shi~\& Krolik (2015) have performed three
dimensional MHD simulations of circumbinary discs, in which accretion
results from the turbulence produced by the magnetorotational
instability.  They set up a disc with an isothermal equation of state
with a sound speed $c_s=0.1$, corresponding to a gas disc scaleheight
at hydrostatic equilibrium $H \sim 0.1 r^{3/2}$.  As pressure forces
in the cavity are very small, the motion of the flow there is
essentially ballistic and therefore particles move radially on the
dynamical timescale.  However, it was found that hydrostatic
equilibrium still had time to be established in the vertical direction
(J.--M.~Shi \& J.~Krolik, private communication), so that the gas disc
scale height in the cavity varies in the same way as in the rest of
the disc.  In that case, as the aspect ratio of discs around pre--main
sequence stars is on the order of 0.1, the value of $H/r=10^{-3}$
needed to fit the lightcurve of CoRoT~223992193 would be that of the
dust but could not be that of the gas, which would imply that dust and
gas were not well mixed in the vertical direction.  In the simulations
performed by Shi \& Krolik (2015), the flow is found to be laminar in
the inner cavity.  This is because the flow moves in too rapidly for
magnetic instabilities to develop into turbulence (J. Krolik, private
communication).  In such a situation, dust grains would be expected to
sedimente toward the disc midplane, which would lead to a dust disc
much thinner than the gas disc (only when turbulence is present can
dust grains diffuse all the way up to the disc surface and be well
mixed with the gas).  In that case, adopting a smaller value of $H/r$
when calculating the light curve than when modelling the disc simply
amounts to ignoring the gas which is above the dust layer.  It is not
clear however that there would be enough time for the grains to
sedimente while they are crossing the cavity.  For the gas densities
used in our modelling, the settling timescale is on the order of
$10^4$--$10^5$~years (e.g., Dullemond \& Dominik 2004).  Therefore, if
the particles are well mixed with the gas before entering the cavity,
they would not have time to settle while flowing through the cavity.
On the other hand, if the flow were not forced to be isothermal but
instead proper cooling were used in the simulations, we may find that
the very low density in the inner cavity enables the gas to cool down
efficiently, thus reducing the scale height significantly, so that
$H/r=10^{-3}$ could represent the scale height of the gas disc as well
as that of the dust disc.  If that were the case, our calculations
would not be self--consistent, as we are using the same fixed value of
$c_s$ in the disc and in the cavity.  However, the overall $r$-- and
$\theta$--dependence of the gas surface density corresponding to a
varying $c_s$ would probably not be very different from what we have
been using, so that our conclusions should still be valid.

As the absorption comes almost entirely from the regions near $\rmin$,
the variations of the light curves we are reproducing should have a
period of about the orbital period at $r=\rmin$, which is 3.2~days.
The light curves displayed in figures~\ref{fig3} and~\ref{fig4} are
rather symmetric about $\theta=180\degr$, and in that case the period
of the variations is actually half the orbital period.  However, we
have made the model symmetric by defining the inner edge of the region
in which there are dust grains as a circle of radius $r=\rmin$.  In
reality, since the sublimation radius is different for the two stars,
the geometry is more complicated and there may not be the symmetry
observed in the calculations.  In addition, the flow inside the cavity
is likely to be very unstable with fluctuations which are not captured
by our model (Shi \& Krolik 2015).

\section*{Acknowledgements}
We thank Ed Gillen and Suzanne Aigrain for very informative
discussions about the variability of CoRoT~223992193.

%
%===========================================================
%

%\bibliographystyle{apj}
%\bibliographystyle{plain}
%\bibliographystyle{mn2e}
%\bibliography{biblio_papers}

\begin{thebibliography}{}

\bibitem[2014]{C14} Cody A. M., {\em et al.}, 2014, AJ, 147, 82

\bibitem[2006]{deValBorro06} de Val--Borro M., Edgar R. G., Artymowicz
  P. {\em et al.}, 2006, MNRAS, 370, 529

\bibitem[2011]{VB11} de Val--Borro M., Gahm G. F., Stempels H. C.,
Peplinski A., 2011, MNRAS, 413, 2679

\bibitem[2014]{G14} Gillen E. {\em et al.}, 2014, A\&A, 62, A50 

\bibitem[2015]{G15} Gillen E. {\em et al.}, 2015, {\em submitted}

\bibitem[2010]{H10} Hanawa T., Ochi Y., Ando K., 2010, ApJ, 708, 485

\bibitem[1997]{JM97} Jensen E. L. N.,  Mathieu R. D., 1997, AJ, 114, 301

\bibitem[1979]{LP79} Lin D. N. C.,  Papaloizou J., 1979, MNRAS, 188, 191

\bibitem[2007]{Mignone07} Mignone A., Bodo G., Massaglia S.,
  Matsakos T., Tesileanu O., Zanni C., Ferrari A., 2007, ApJS, 170, 228

\bibitem[1977]{P77} Paczynski B., 1977, ApJ, 216, 822

\bibitem[1977]{PP77} Papaloizou J.,  Pringle J. E., 1977, MNRAS, 181, 441

\bibitem[2012]{Shi12} Shi J.--M., Krolik J. H., Lubow S. H., Hawley
  J. F., 2012, ApJ, 749, 118

\bibitem[2015]{Shi15} Shi J.--M., Krolik J. H., 2015,  ApJ, 807, 131

\bibitem[2014]{S14} Stassun K. G., Feiden G. A., Torres G. 2014,
  NewAR, 60, 1

%\bibitem[2014]{Bucchave} Buchhave L. A., Bizzarro M., Latham D. W.
%  {\em et al.}, 2014, arXiv:1405.7695

%\bibitem[1968]{Cox}
%Cox J. P., Giuli R. T., 1968, Principles of Stellar
%Structure: Physical Principles (New York: Gordon \& Breach)

\end{thebibliography}

%
%===========================================================

\label{lastpage}
\end{document}